\documentclass[aap,MSNbibl,citesort,dvips]{arximspdf}

\doi{10.1214/11-AAP804}
\volume{22}
\issue{4}
\pubyear{2012}
\firstpage{1642}
\lastpage{1649}

\makeatletter

\newtheorem{theorem}{Theorem}[section]
\newtheorem{lemma}[theorem]{Lemma}
\newtheorem{claim}[theorem]{Claim}

\makeatother

\begin{document}
\begin{frontmatter}

\title{Balanced allocation: Memory performance~tradeoffs}
\runtitle{Balanced allocation: Memory performance tradeoffs}

\begin{aug}
\author[A]{\fnms{Itai} \snm{Benjamini}\ead[label=e1]{itai.benjamini@weizmann.ac.il}}
\and
\author[B]{\fnms{Yury} \snm{Makarychev}\corref{}\ead[label=e2]{yury@ttic.edu}}
\runauthor{I. Benjamini and Y. Makarychev}
\affiliation{Weizmann Institute and Toyota Technological Institute at Chicago}
\address[A]{Department of Mathematics\\
Weizmann Institute\\
Rehovot 76100\\
Israel\\
\printead{e1}}
\address[B]{Toyota Technological Institute\\
\quad at Chicago\\
6045 S. Kenwood Ave.\\
Chicago, Illinois 60637\\
USA\\
\printead{e2}} 
\end{aug}

\received{\smonth{11} \syear{2009}}
\revised{\smonth{2} \syear{2011}}

%
\begin{abstract}
Suppose we sequentially put $n$ balls into $n$ bins. If we put each
ball into a random bin then the heaviest bin will contain ${\sim} \log
n /\log\log n$ balls with high probability. However, Azar, Broder,
Karlin and Upfal [\textit{SIAM J. Comput.} \textbf{29} (1999)
180--200] showed that if each time we choose two bins at random and
put the ball in the least loaded bin among the two, then the heaviest
bin will contain only ${\sim}\log\log n$ balls with high probability.
How much memory do we need to implement this scheme? We need roughly
$\log\log\log n$ bits per bin, and $n\log\log\log n$ bits in total.

Let us assume now that we have limited amount of memory. For each ball,
we are given two random bins and we have to put the ball into one of
them. Our goal is to minimize the load of the heaviest bin. We prove
that if we have $n^{1-\delta}$ bits then the heaviest bin will contain
at least $\Omega(\delta\log n/\log\log n)$ balls with high
probability. The bound is tight in the communication complexity model.
\end{abstract}

%

\end{frontmatter}

\section{Introduction}
Suppose we sequentially put $n$ balls into $n$ bins.
If we put each ball in a bin chosen independently and uniformly at random,
the maximum load (the largest number of balls in any bin)
will be ${\sim}\log n /\log\log n$ with high probability.
We can significantly reduce the maximum load by using the ``power of
two choices''
scheme of Azar, Broder, Karlin and Upfal~\cite{ABKU}: if we put
each ball in the least loaded of two bins chosen independently and
uniformly at random, the maximum load will be ${\sim}\log\log n$ with
high probability.
This scheme has numerous applications for hashing, server load
balancing and
low-congestion circuit routing
(see~\cite{ACMR,ABKU,BCSV,BCM1,BCM2,CMMMRSSV,MRR,TW}).

As an example, consider an implementation of a hash table that uses the
``power of two choices'' paradigm. We keep a table of size $n$;
each table entry can store multiple elements (say) in a doubly-linked list.
We use two perfectly random hash functions $h_1$ and $h_2$ that map
elements to table entries. To insert an element $e$, we find two possible
table entries $h_1(e)$ and $h_2(e)$, and store the element in
the table entry with fewer elements. To find an element $e$,
we search through all elements in entries $h_1(e)$ and $h_2(e)$.
This requires only $O(\log\log n)$ operations for
every element $e$ w.h.p.; whereas if we used only one hash function
we would need to perform $\Omega(\log n/\log\log n)$ operations for some
elements w.h.p.

How many extra bits of memory do we need to implement this scheme?
We need roughly $\log\log\log n$ bits per bin (table entry) to store
the number of balls (elements) in the bin, and $n\log\log\log n$ bits
in total.

Let us assume now that we have limited amount of memory. For each ball,
we are given two random bins (e.g., we are given two hash values)
and we have to put the ball into one of them. Can we still
guarantee that the maximum load is $O(\log\log n)$ with high probability?

The correct answer is not obvious. One could assume that if the number
of memory bits is $o(n)$ then
the maximum load should be ${\sim}\frac{\log n}{\log\log n}$
balls. However, that is not the case as the following example shows.
Let us group all bins into $n/\log\log n$ clusters; each cluster
consists of $\log\log n$ bins. For each cluster, we keep the total
number of balls in the bins that form the cluster. Now given a ball
and two bins, we put the ball into the bin whose cluster contains
fewer balls. The result of Azar, Broder, Karlin and Upfal~\cite{ABKU}
implies that w.h.p. each cluster will contain at most
$\frac{n}{n/\log\log n} + \log\log n = 2\log\log n$ balls. Therefore,
each bin will also contain at most $2\log\log n$ balls.
This scheme uses $\frac{n}{\log\log n} \log\log\log n = o(n)$ bits
of memory.

In this paper, we show that if we have $n^{1-\delta}$
bits of memory then the maximum load is
$\Omega(\delta\log n/\log\log n)$ balls with high probability.
%
We study the problem in the ``communication complexity model.''
In this model, the state of the algorithm is determined by $M$ bits of memory.
Before each step, we choose the memory state $m \in\{1,\ldots, 2^M\}$.
Then the algorithm gets two bin choices $i$ and $j$.
It selects one of them based on $m$, $i$, $j$ and 
independent random bits.
That is, the algorithm chooses $i$ with a certain probability
$f(m, i, j)$ and $j$
with probability $1 - f(m, i, j)$; the choice is independent from the
previous steps.

Unlike the standard computational model, we do not require that the
memory state of the algorithm
depends only on $m$, $i$, $j$ and the random bits in the communication
complexity model.
In particular, the state can depend on the current load of bins.
Hence, algorithms in our model are more powerful than algorithms in the
computational model.
Consequently, our lower bound (Theorem~\ref{thmmain}) applies also to
the computational model,
whereas our upper bound (Theorem~\ref{thmtight}) applies only to the
communication complexity model.

First, we prove the lower bound on maximum load.
%
\begin{theorem}\label{thmmain}
We are sequentially given $n$ balls. We have to put
each of them into one of two bins chosen uniformly and independently at
random among $n$ bins. We have only
$M = n^{1-\delta}$ bits of memory $(\delta> 0$ may depend on $n)$;
our choice where to put a ball can depend only on these memory bits and
random bits.
Then the maximum load will be at least $\frac{\delta\log n}{2\log
\log n}$ with probability $1 - o(1)$.
\end{theorem}

Then we show that the bound is essentially tight in the communication
complexity model.
%
\begin{theorem}\label{thmtight}
There exists an algorithm that gets $M = n^{1-\delta}$ bits of advice
before each step and uses
no other memory, and ensures that the heaviest bin contains at most
$O(\frac{\delta\log n}{\log\log n})$ balls w.h.p.
$[$where $\delta\geq1/(\log n)^{1-\Omega(1)}]$.
\end{theorem}

In Section~\ref{secmain}, we prove Theorem~\ref{thmmain}. In
Section~\ref{sectight},
we prove Theorem~\ref{thmtight}.

\section{\texorpdfstring{Proof of Theorem \protect\ref{thmmain}}{Proof of Theorem 1.1}}
\label{secmain}
We assume that $\frac{\delta\log n}{2\log\log n} \geq1$, as
otherwise the statement of the theorem
is trivially true (there is a bin that contains at least one ball).

Consider one step of the bins--and--balls process: we are given two bins
chosen uniformly at random,
and we put the ball into one of them.
Let $p_i\equiv p_i^{(m)}$ be the probability that we put the ball into
bin $i$ given
that the memory state is $m\in\{1,\ldots, 2^M\}$. Let $F_m \equiv
F^{\varepsilon}_m = \{i\dvtx p^m_i < \varepsilon/n\}$.
%
\begin{claim}
\label{claimmain}
(1) For every set of bins $S$, the probability that we put a~ball in a
bin from $S$ is at least
$\varepsilon|S\setminus F_m^\varepsilon| / n$:
\[
\mbox{(2)}\quad|F_m^\varepsilon| \leq\varepsilon n.
\]
\end{claim}
\begin{pf}
(1) The desired probability equals
\[
\sum_{i\in S} p_i \geq\sum_{i\in S\setminus F_m^\varepsilon} p_i
\geq\frac{\varepsilon
|S\setminus F^\varepsilon_m|}{n}.
\]

(2) The probability that both chosen bins are in $F^\varepsilon_m$ is
$|F^\varepsilon_m|^2/n^2$. Therefore, the probability $t$
that we put the ball into a bin from $F^\varepsilon_m$ is at least
$|F^\varepsilon_m|^2/n^2$.
On the other hand, we have
\[
t = \sum_{i\in F^\varepsilon_m} p_i < \frac{\varepsilon
|F^\varepsilon_m|}{n}.
\]
We conclude that $|F^\varepsilon_m| \leq\varepsilon n$.
\end{pf}

We divide the process into $L$ consecutive phases. In each phase, we
put $\lfloor n/L\rfloor$ balls into bins. Let $S_i$ be the set of bins
that contain at least $i$
balls at the end of the phase $i$; let $S_0 = \{1,\ldots, n\}$. Now we
will prove
a bound on the size of $S_i$ that in turn will imply Theorem~\ref{thmmain}.\vadjust{\goodbreak}
%
\begin{lemma} Let $L = \lceil\frac{\delta}{2} \log n/ \log\log
n\rceil$, $\varepsilon= 1/(2L)$ and
$\beta= 1/(4L)$. For every $i\in\{0,\ldots,L\}$, let ${\mathcal E}_i$
be the event that
for every $m_1,\ldots,m_{L-i} \in\{1,\ldots,2^M\}$,
\[
\Biggl|S_i\biggm\backslash\bigcup_{j=1}^{L-i} F_{m_j}\Biggr|
\geq\frac{(\beta\varepsilon)^i}{2} n.
\]
Then for every $i$
\[
\Pr({\mathcal E}_i) =1 - o(1).
\]
In particular, $\Pr(|S_L| > 0) \geq\Pr({\mathcal E}_L) = 1 - o(1)$,
and therefore, in the end, the heaviest
bin contains at least $L$ balls w.h.p.
\end{lemma}
\begin{pf}
First, note that the event ${\mathcal E}_0$ always holds,
\[
\Biggl|S_0\biggm\backslash\bigcup_{j=1}^{L} F_{m_j}\Biggr| \geq n -
L\varepsilon n = n/2.
\]

Now we shall prove that $\Pr(\bar{{\mathcal E}_i}|{\mathcal E}_{i-1})
\leq o(1/L)$ (uniformly for all $i$), and thus
\begin{eqnarray*}
\Pr({\mathcal E}_i) &\geq& \Pr({\mathcal E}_0\wedge\cdots\wedge
{\mathcal E}_i) =
1 - \sum_{j=1}^i \Pr({\mathcal E}_{0} \wedge\cdots\wedge{\mathcal
E}_{j-1} \wedge\bar{{\mathcal E}_j})
- \Pr(\bar{{\mathcal E}_0})\\
&\geq& 1 - \sum_{j=1}^i \Pr(\bar{{\mathcal E}_j} \wedge{\mathcal
E}_{j-1}) \geq
1 - \sum_{j=1}^i \Pr(\bar{{\mathcal E}_j} | {\mathcal E}_{j-1}) = 1
- o(1).
\end{eqnarray*}

Assume that ${\mathcal E}_{i-1}$ holds.
Fix $m_1,\ldots, m_{L-i}$.
We are going to estimate the number of bins in $S_{i-1}\setminus
\bigcup_{j=1}^{L-i} F_{m_j}$ which we put a ball into during the phase~$i$.
All those bins are in the set $S_i\setminus\bigcup_{j=1}^{L-i} F_{m_j}$.

Consider one step of the process; we are given the $t$th ball (in the
current phase)
and have to put it in a bin. Let $N_{t-1}$ be the set of bins in
$S_{i-1}\setminus\bigcup_{j=1}^{L-i} F_{m_j}$ where we have already
put a ball into
(during the current phase). We are going to lower bound the probability
of the event that
we put the ball into a ``new bin,'' that is, in a bin in
$S_{i-1}\setminus\bigcup_{j=1}^{L-i} F_{m_j}\setminus N_{t-1}$.
Denote the indicator variable of this event by $q_t$.
Let $m$ be the state of the memory at time $t$. Since ${\mathcal
E}_{i-1}$ holds,
\[
\Biggl|\Biggl(S_{i-1}\biggm\backslash\bigcup_{j=1}^{L-i} F_{m_j}
\Biggr)\biggm\backslash F_{m}\Biggr| \geq\frac{(\beta\varepsilon)^{i-1}n}{2}.
\]
Therefore, by Claim~\ref{claimmain}(1), the probability that $q_t=1$
is at least
\[
\frac{\varepsilon|S_{i-1}\setminus\bigcup_{j=1}^{L-i}
F_{m_j}\setminus F_m
\setminus N_{t-1}|}{n} \geq\biggl(\frac{(\beta\varepsilon
)^{i-1}}{2} - \frac{|N_{t-1}|}{n}\biggr)\varepsilon.
\]
Thus, if $|N_{t-1}| \leq(\beta\varepsilon)^{i-1} n/4$,
\[
\Pr(q_t=1 | q_1,\ldots, q_{t-1}) \geq(\beta\varepsilon)^{i-1}
\times\varepsilon/4
\stackrel{\mathrm{def}}{=} \mu.
\]
Note that $|N_t| = |N_{t-1}| + q_t$ and $|N_t| = q_1 + \cdots+ q_t$.
Now we want to apply the Chernoff bound to the random variables $\{q_j\}
_j$. However, since they
are not necessarily independent, we will need an additional step.
Define random variables $\tilde q_j$ as follows.

If $|N_{t-1}| \leq(\beta\varepsilon)^{i-1} n/4$,
\[
\cases{\mbox{if } q_t = 1, &\quad let $\tilde q_t = 1 \mbox{ w.p. } \dfrac
{\mu}{\Pr(q_t=1|q_1,\ldots,
q_{t-1})}$;\vspace*{2pt}\cr
\mbox{if } q_t = 1, &\quad let $\tilde q_t = 0 \mbox{ w.p. } 1 - \dfrac{\mu
}{\Pr(q_t=1|q_1,\ldots,
q_{t-1})}$;\vspace*{2pt}\cr
\mbox{if } q_t = 0, &\quad let $\tilde q_t = 0$.}
\]
If $|N_{t-1}| > (\beta\varepsilon)^{i-1} n/4$,
\[
\cases{
\mbox{let } \tilde q_t = 1 \mbox{ w.p. } \mu;\cr
\mbox{let } \tilde q_t = 0 \mbox{ w.p. } 1 - \mu.}
\]
It is easy to see that in either case
$\Pr(\tilde q_t= 1 | \tilde q_1, \ldots, \tilde q_{t-1}) = \mu$. Therefore,
$\tilde q_1,\ldots, \tilde q_t$ are i.i.d. 0--1 Bernoulli random
variables with expectation $\mu$.
By the Chernoff bound, the probability that $\tilde q_1 +\cdots+
\tilde q_{n/L}$ is at least
\[
\frac{1}{2} \times\mathbb{E}[\tilde q_1 +\cdots+ \tilde q_{n/L}] =
\frac12 \times\frac{n\mu}{L} = \frac{(\beta\varepsilon)^{i} n}{2}
\]
is at least
$1 - 2\cdot2^{-(\beta\varepsilon)^i n/8}$.
Since $q_t \geq\tilde q_t$ if $|N_{t-1}| < (\beta\varepsilon)^{i-1} n/4$,
\[
|N_t| = q_1 + \cdots + q_t \geq\min\bigl((\beta\varepsilon)^{i-1} n/4,
\tilde q_1 + \cdots + \tilde q_t\bigr).
\]
Finally, we have
\begin{eqnarray*}
\Pr\biggl(|N_{n/L}| \geq\frac{(\beta\varepsilon)^{i} n}{2}\biggr)
&\geq&
\Pr\biggl(\min\bigl((\beta\varepsilon)^{i-1} n/4, \tilde q_1 + \cdots
+\tilde q_{n/L}\bigr)
\geq\frac{(\beta\varepsilon)^{i} n}{2}\biggr) \\
&=&
\Pr\biggl(\tilde q_1 + \cdots +\tilde q_{n/L} \geq\frac{(\beta
\varepsilon)^{i} n}{2}\biggr)
\geq1 - 2\cdot2^{-(\beta\varepsilon)^i n/8}.
\end{eqnarray*}
Since $S_{i} \setminus\bigcup_{j=1}^{L-i} F_{m_j} \supset N_{n/L}$,
\[
\Pr\Biggl(\Biggl|S_i \biggm\backslash\bigcup_{j=1}^{L-i} F_{m_j} \Biggr| \geq\frac
{(\beta\varepsilon)^i}{2}n
\Big| {\mathcal E}_{i-1}\Biggr) \geq1 - 2\cdot2^{-
(\beta\varepsilon)^i n/8}
\]
for fixed $m_1,\ldots, m_{L-i}$.
By the union bound [recall that $\varepsilon= 1/(2L)$ and $\beta= 1/ (4L)$]
%
\begin{eqnarray}\label{equnion1}
&&\Pr\Biggl(\mbox{for all } m_1,\ldots, m_{L-i}\dvtx\Biggl|S_i \biggm\backslash
\bigcup_{j=1}^{L-i} F_{m_j}\Biggr|
\geq(\beta\varepsilon)^i n/2 \Big| {\mathcal E}_{i-1}\Biggr)
\nonumber\\
&&\qquad\geq1 - 2\cdot(2^{M})^{L-i} 2^{- (\varepsilon\beta)^i n/8}
\geq1- 2^{ML -  ({1}/({8L^2}))^{L} n /8} \\
&&\qquad=
1- 2^{n^{1-\delta}L (1 - n^{\delta}L (
{1}/({8L^2}))^{L+1})}.\nonumber
\end{eqnarray}
Recall that $L = \lceil\frac{\delta\log n }{2\log\log n} \rceil$.
We have,
\begin{eqnarray*}
(8L^2)^{L+1} &\leq& (8L^2)^2 \cdot(8L^2)^{\delta\log
n/(2\log\log n)}\leq
(8L^2)^2 \cdot\biggl(\frac{\log n}{\omega(1)}\biggr)^{\delta\log
n/\log\log n}\\
&\leq& (8L^2)^2 n^{\delta} 2^{-\omega(L)} =
n^{\delta} 2^{6 + 4 \log L -\omega(L)} = o(n^{\delta}).
\end{eqnarray*}
Therefore, expression (\ref{equnion1}) is $1 - 2^{n^{1-\delta}L (1 -
\omega(L))} = 1 - o(1)$.
\end{pf}

\section{\texorpdfstring{Proof of Theorem \protect\ref{thmtight}}{Proof of Theorem 1.2}}\label{sectight}
In this section, we will prove that our bound is tight in the communication
complexity model. Specifically, we present an algorithm that gets
$M = n^{1-\delta}$ bits of advice before each ball is thrown,
and ensures that the maximum load is at most $O(\frac{\delta
\log n}{\log\log n})$ w.h.p.
when $\delta\geq1/(\log n)^{1-\Omega(1)}$.

Observe that no matter which of the two bins we choose at each step,
the probability $p_i$ that we put the ball in a bin $i$ is at most $2/n$.
Therefore, the probability that after $n$ steps the total number of balls
in the bin $i$ exceeds $T = \frac{2\delta\log n}{\log\log n}
(1+\frac{2\log(1/\delta)}{\log(\delta\log n)})$
is asymptotically at most the probability that a Poisson random
variable with $\lambda=2$ exceeds $T$,
that is, it is at most $e^{-2}\frac{2^T}{T!} (1+ o(1))= o((\frac
{2e}{T})^T) = o(1/(n^\delta\log n))$.
Thus the number of bins that contain at least $T$ balls is at most
$n^{1 - \delta} / (2\log n)$ w.h.p. Before each step, our algorithm
receives the list $L$
of such bins, and the number of balls in each of them. Now if one of
the two randomly
chosen bins belongs to $L$ and the other does not, the algorithm puts
the ball into
the bin that is not in~$L$; if both bins are in~$L$, the algorithm puts
the ball into the bin with fewer balls (let us say that we use the
``always-go-left''
tie breaking rule: if both bins contain the same number of balls,
we put the ball into the left of the two bins); finally, if both bins
are not in $L$,
the algorithm puts the ball into an arbitrary bin.

Let us estimate the maximum load. We say that a ball is an ``extra
ball'' if we
put it into a bin that is in $L$ (at the moment when we put the ball).
Then the total number of balls in a bin is at most $T$ plus the number
of extra balls in the bin. Let us now count only extra balls. Note that
every time
we get a ball, we either:
\begin{itemize}
\item``discard it,'' put it into a bin that is not in $L$, and thus do
not count it as an extra ball, or
\item put it into one of the two bins that contains fewer ``extra balls.''
\end{itemize}
That is, we use a modified scheme of Azar, Broder, Karlin and Upfal, where
we sometimes put a ball into one of the two bins that contains fewer
``extra balls,''\vadjust{\goodbreak}
and sometimes discard it. We claim that each bin contains at most $\log
\log n$ extra balls
as in the standard ``power of two choices'' scheme of Azar, Broder,
Karlin and Upfal.
%
\begin{claim}
Consider the balls and bins process. Suppose at step $i$ we are given
the choice of two bins $a_i^1$ and $a_i^2$. Let $k_{ij}$ be the number of
balls in bin $j$ after $i$ steps when we use the standard ``power of
two choices''
scheme. Let $\tilde k_{ij}$ be the number of extra balls in bin $j$
after $i$ steps when we use
our modified ``power of two choices'' scheme.
Assume that in both cases we use the ``always-go-left'' tie breaking rule.
Then 
$\tilde k_{ij} \leq k_{ij}$,
for every $1\leq i,j \leq n$ $($the statement holds for \textit{every}
sequence $\{a_i^1,a_i^2\}_{i=1,\ldots, n})$.
\end{claim}
\begin{pf}
We prove that $\tilde k_{ij} \leq k_{ij}$ by induction on $i$.
Initially, all
bins contain no balls, $\tilde k_{0j} = k_{0j} = 0$, so the statement holds.
Assume that the statement holds for $i < i_0$, we verify that
$\tilde k_{ij} \leq k_{ij}$ for $i=i_0$. Fix $j$.
Consider several cases.
\begin{itemize}
\item First, suppose that we put the ball into the bin $j$ at step $i$
in both schemes.
Then $\tilde k_{ij} = \tilde k_{i-1,j} + 1 \leq k_{i-1,j} + 1 = k_{ij}$.
\item Now suppose that we put the ball into the bin $j$ at step $i$ in
the modified scheme, however,
we put the ball into some bin $j'\neq j$ in the standard scheme.
Note that if $j < j'$ then $\tilde k_{i-1,j} \leq\tilde k_{i-1,j'}$
and $k_{i-1,j'} < k_{i-1,j}$ thus
\[
\tilde k_{ij} = \tilde k_{i-1,j} + 1 \leq\tilde k_{i-1,j'} + 1 \leq
k_{i-1,j'} + 1 \leq k_{i-1,j}
=k_{ij};
\]
if $j' < j$ then
$\tilde k_{i-1,j} < \tilde k_{i-1,j'}$
and $k_{i-1,j'} \leq k_{i-1,j}$ and thus
\[
\tilde k_{ij} = \tilde k_{i-1,j} + 1 \leq\tilde k_{i-1,j'} \leq
k_{i-1,j'}\leq k_{i-1,j} = k_{ij}.
\]
\item Finally, suppose that in the modified scheme we put the ball into
some bin $j'\neq j$ or discard it at step $i$.
Then $\tilde k_{ij} = \tilde k_{i-1,j}\leq
k_{i-1,j} \leq k_{ij}$.\qed
\end{itemize}
\noqed\end{pf}

Note that if bins $a_i^1$ and $a_i^2$ are chosen uniformly at random, then
$\max_j k_{nj} = \log\log n + \Theta(1)$ with high probability~\cite{ABKU}.
Therefore, by the claim, $\max_j \tilde k_{nj} = \log\log n + O(1)$,
and each bin contains at most $T + \log\log n + O(1) = O(\frac
{\delta\log n}{\log\log n})$
balls w.h.p.

\section*{Acknowledgments}

Noga Alon showed us the no memory case before pursuing this work. We
would like to thank Noga Alon and Eyal Lubetzky for useful discussions.
We thank the anonymous referee for valuable suggestions.



\printaddresses

\end{document}